\documentclass[twocolumn,prd,amsmath,superscriptaddress,nofootinbib,floatfix]{revtex4}

\usepackage{amsmath}
\usepackage{bm}
\usepackage{graphicx}
\usepackage{color}
\usepackage{subfigure}

\def\gsim{\;\rlap{\lower 2.5pt
 \hbox{$\sim$}}\raise 1.5pt\hbox{$>$}\;}
\def\lsim{\;\rlap{\lower 2.5pt
 \hbox{$\sim$}}\raise 1.5pt\hbox{$<$}\;}

\def\be{\begin{equation}}
\def\ee{\end{equation}}
\def\bea{\begin{eqnarray}}
\def\eea{\end{eqnarray}}

\begin{document}

\title{Anisotropic Extinction Distortion of the Galaxy Correlation Function}

\author{Wenjuan Fang}
\affiliation{Department of Physics, University of Michigan, Ann Arbor, MI 48109}
\affiliation{Department of Physics, Columbia University, New York, NY 10027}
\author{Lam Hui}
\affiliation{Department of Physics, Columbia University, New York, NY 10027}
\affiliation{Institute for Strings, Cosmology and Astroparticle Physics, Columbia University, New York, NY 10027}
\author{Brice M$\acute{\rm e}$nard}
\affiliation{Canadian Institute for Theoretical Astrophysics, Toronto, Ontario, M5S 3H8, Canada}
\affiliation{Department of Physics $\&$ Astronomy, Johns Hopkins University, 366 Bloomberg Center, 3400 N. Charles Street, Baltimore, MD 21218}
\author{Morgan May}
\affiliation{Brookhaven National Laboratory, Upton, NY 11973}
\author{Ryan Scranton}
\affiliation{Department of Physics, University of California, Davis, CA, 95616}
\date{\today}

\begin{abstract}

Similar to the magnification of the galaxies' fluxes by gravitational lensing, the extinction of the fluxes by comic dust, whose existence is recently detected by \cite{MenSFR09}, also modify the distribution of a flux-selected galaxy sample. We study the anisotropic distortion by dust extinction to the 3D galaxy correlation function, including magnification bias and redshift distortion at the same time. We find the extinction distortion is most significant along the line of sight and at large separations, similar to that by magnification bias. The correction from dust extinction is negative except at sufficiently large transverse separations, which is almost always opposite to that from magnification bias (we consider a number count slope $s > 0.4$). Hence, the distortions from these two effects tend to reduce each other.
At low $z$ ($\lsim 1$), the distortion by extinction is stronger than that by magnification bias, but at high $z$,
the reverse holds.
We also study how dust extinction affects probes in real space of the baryon acoustic oscillations (BAO) and the linear redshift distortion parameter $\beta$. We find its effect on BAO is negligible. However, it introduces a positive scale-dependent correction to $\beta$ that can be as large as a few percent. At the same time, we also find a negative scale-dependent correction from magnification bias, which is up to percent level at low $z$, but to $\sim 40\%$ at high $z$. These corrections are non-negligible for precision cosmology, and should be considered when testing General Relativity through the scale-dependence of $\beta$.

\end{abstract}

\maketitle

\section{Introduction}

Galaxy redshift surveys provide a 3-dimensional map of the universe's large scale structure (LSS), and hence serve as important probes of cosmology. The simplest statistic obtained from a redshift survey is the two point galaxy correlation function, or its Fourier transform, the galaxy power spectrum. Both quantities have been measured to good accuracy by recent large redshift surveys such as the 2dF Galaxy Redshift Survey (2dFGRS) \cite{Col01,Col03} and the Sloan Digital Sky Survey (SDSS) \cite{Yor00}, and the results have been used to put constraints on cosmological parameters, e.g., \cite{Eis05,Oku08,CabG09,San09,Kaz10a,Chu10} by using the correlation function and \cite{Col05,Hue06a,Hue06b,Teg06,Per07a,Per07b,Per07c,Per10,Rei10} by using the power spectrum.

It is known that the measurements of these two quantities are subject to several effects, which distort them to be anisotropic, i.e. the measured signal depends on the orientation
of the separation or wave vector.
One effect comes from the possibly wrong cosmology used to convert the measured coordinates, i.e. redshifts and angular positions, to the comoving ones---the Alcock-Paczynski effect (\cite{AlcPac79,MatS96,Bal96,HSB99,HuHai03}). A second one comes from the peculiar velocities of the galaxies that introduce uncertainties in the interpretation from redshift to comoving distance---the redshift distortion (\cite{Dav83,Kaiser87,Ham92,S04}). A third effect is caused by gravitational lensing mainly through magnification of the galaxies' fluxes (for flux-selected galaxy samples) and
changes to the apparent angular separations between galaxies---the magnification bias \cite{Tur84,WHHW88,N89,G03,SM05,Mat00,HuiGL07a,HuiGL07b}.

For measurements made with galaxy samples that are selected by the galaxy's flux, a fourth anisotropic effect would arise from the existence of cosmic dust, which causes extinction in the fluxes of the galaxies.
By cosmic dust we mean dust which is correlated with galaxies, but which may or may not reside
in galaxy haloes. In a flux selected galaxy sample, such cosmic dust would
modulate the galaxy density field -- fewer galaxies behind regions of higher extinction.
Moreover, the effect is anisotropic in 3D. Recall that a two-point correlation measurement
is essentially a measurement of pair counts. At a given separation,
pairs of galaxies that are aligned close to
the line of sight suffer more of an extinction effect -- i.e. dust correlated with
the foreground galaxies dims (and removes from sample) the background galaxies --
while pairs of galaxies oriented transverse to the line of sight are less susceptible.
This is much like how gravitational lensing or magnification bias introduces
an anisotropy to the galaxy correlation function or power spectrum \cite{Mat00,HuiGL07a,HuiGL07b}.
The difference is that gravitational lensing by the foreground galaxies
generally brightens the background galaxies, thus adding them to one's sample,
and it also causes an overall geometrical stretching which dilutes the apparent number density.
As we will see, the precise shapes of these two anisotropies are different. We also notice here that there could be other anisotropic effects, e.g. if the selection is orientation-dependent, anisotropy would be introduced for galaxies that are aligned by the large scale tidal fields~\cite{H09}.

From the observational side, the existence of cosmic dust has recently been detected by \cite{MenSFR09}: by using a quasar sample at $z>1$ and a galaxy sample at $z\sim0.3$ from the SDSS, the authors find a positive correlation between the redness of the background quasars and the overdensity of the foreground galaxies, up to an angular separation of $\sim 2^{\circ}$ or, a corresponding projected distance separation of $\sim20h^{-1}$Mpc at $z\sim0.3$, which indicates the existence of dust correlated with galactic
haloes and the LSS.
From the brightness of these quasars, the authors also find that the extinction by cosmic dust occurs at a level comparable to the magnification by gravitational lensing. Hence, it is
important to take into account of the effect of dust extinction and evaluate its significance for studies of LSS.

In this paper, by using the results from \cite{MenSFR09}, we investigate the effect of dust extinction on the galaxy correlation function, focusing on the anisotropic features it induces, and study how this affects cosmological probes through measurements of the galaxy correlation function such as the Baryon Acoustic Oscillations (BAO) and the linear redshift distortion, e.g. \cite{wigglez,boss,bigboss,euclid}. Our calculation accounts for distortion from
extinction, as well as that from peculiar motion and lensing.

The outline of the paper is as follows. In \S \ref{sec:calculation} we derive the formulas used for our calculation, with technical details relegated to the Appendices. In \S \ref{sec:anisotropy}, we present and discuss the anisotropic features caused by dust extinction, focusing on a comparison with those by gravitational lensing. The effects of dust extinction on cosmological probes such as BAO and linear redshift distortion are studied in \S \ref{sec:systematic}. Finally, we conclude in \S \ref{sec:discuss}.

\section{Calculation of the Distortion}
\label{sec:calculation}

\subsection{Fluctuation of Dust Extinction}

The optical depth due to dust extinction to a source at a comoving distance
$\chi$ away and angular position $\boldsymbol{\theta}$,
and at observed wavelength $\lambda_{\rm obs}$, takes the form:
\begin{eqnarray}
\tau(\chi, \boldsymbol{\theta}; \lambda_{\rm obs})
= \int_{0}^{\chi}\frac{d\chi'}{(1+z')} {\rho}_{\rm d} (\chi',\boldsymbol{\theta}) f(\chi', \lambda_{\rm obs}) \, ,
\end{eqnarray}
where $\rho_{\rm d}$ is the proper mass density of dust,
and $z'$ is the redshift associated with the comoving radial distance $\chi'$ on the light cone, (hereafter, we would use $z$, $\chi$ interchangeably.)
Here $f$ is the extinction efficiency --
more precisely, $\rho_d f$ gives the inverse (proper) mean free path of photons scattered by dust.
For simplicity, we assume the intrinsic dust extinction properties do not fluctuate
spatially -- namely, $f$ is a constant at fixed rest-frame wavelength.
Thus fluctuations in the optical depth arise entirely from fluctuations in dust density.
Subtracting the mean optical depth from the above expression, we have
\begin{eqnarray}
\delta\tau(\chi,\boldsymbol{\theta};\lambda_{\rm obs})=\int_{0}^{\chi}\frac{d\chi'}{(1+z')}\bar{\rho}_{\rm d}(\chi')\delta_{\rm d}(\chi',\boldsymbol{\theta}) f(\chi',\lambda_{\rm obs}),
\end{eqnarray}
where $\bar\rho_{\rm d}(\chi')$ is the mean dust density at $\chi'$ and
$\delta_d(\chi', \boldsymbol{\theta})$ is the fractional overdensity of dust. We will discuss how the evolution of $\bar\rho_{\rm d}$ is modeled below.
See \cite{Cor06} for a more detailed discussion.

\subsection{Corrections to the Correlation Function}
\label{subsec:xiobs}

When the effects from peculiar velocities, gravitational lensing and dust extinction are all taken into account, we find that, to first order in perturbations, the observed galaxy overdensity $\delta_{\rm obs}$ is related to its intrinsic counterpart $\delta_g$ by the following
\begin{equation}
\delta_{\rm obs}=\delta_g+\delta_v+\delta_{\mu}+\delta_e,\label{eqn:deltaobs}
\end{equation}
where $\delta_v$, $\delta_{\mu}$, and $\delta_e$ are corrections from the peculiar velocity, gravitational lensing and dust extinction respectively, and they are given by
\begin{eqnarray}
\delta_v &=& -\frac{(1+z)}{H(z)}\frac{\partial v_{\parallel}}{\partial \chi},\\
\delta_{\mu}&=&[5s(z)-2]\kappa,\label{eqn:deltamu}\\
\delta_e&=&-2.5s(z)\delta\tau\label{eqn:deltae},
\end{eqnarray}
where $v_{\parallel}$ is the line-of-sight peculiar velocity, positive if pointing away from us, $H(z)$ is the Hubble expansion rate at the redshift of observation $z$, $\kappa$ is the lensing convergence, $\delta\tau$ is, as before, the fluctuation of the extinction optical depth, $s(z)$ is the factor that describes how the distribution of a galaxy sample is modified by the changes in the individuals' fluxes. Note the two terms containing $s(z)$ in Eqn~(\ref{eqn:deltamu}) and Eqn~(\ref{eqn:deltae}) have opposite signs, which just indicates the opposite effects from dust extinction and gravitational magnification.
For a sharp faint-end cutoff selection of the galaxy sample, $s(z)$ (henceforth
referred to as the number count slope) is given by
\begin{equation}
s(z)=\frac{d\log_{10}\bar{n}_{\rm obs}(z, <m)}{dm}\bigg|_{m=m_{\rm max}}\label{eqn:ngslope},
\end{equation}
where $\bar{n}_{\rm obs}(z, <m)$ is the observed mean of the number density of the galaxies that are brighter than magnitude $m$, and $m_{\rm max}$ is the limiting magnitude for the sample. For a more general sample selection, the expression for $s(z)$ is given in Appendix A, where details of our derivation for the above results are presented. We have suppressed the position dependence of $(\chi,\boldsymbol{\theta})$ on both sides of Eqn~(\ref{eqn:deltaobs}), also for $v_{\parallel}$, $\kappa$, and $\delta\tau$, and the wavelength dependence of $\lambda_{\rm obs}$, the characteristic wavelength of the filter, for $\delta\tau$. Note, in our derivation of $\delta_v$, we have adopted the distant observer approximation and restricted ourselves to sub-horizon scales \cite{Kaiser87, MatS96,Dodelson}. Throughout this paper, we would assume the universe is flat, and set the speed of light $c=1$.

The observed galaxy correlation function $\xi_{\rm obs}(\chi_1,\boldsymbol{\theta_1};\chi_2,\boldsymbol{\theta_2})\equiv$$ \langle \delta_{\rm obs}(\chi_1,\boldsymbol{\theta_1})\delta_{\rm obs}(\chi_2,\boldsymbol{\theta_2}) \rangle$ is then given as a sum of 16 terms
\begin{equation}
\xi_{\rm obs}(1;2)=\sum_{a,b}\xi_{ab}(1;2),
\end{equation}
where we use $1, 2$ as shorthands for $(\chi_1,\boldsymbol{\theta_1})$, $(\chi_2,\boldsymbol{\theta_2})$, ``$a, b$" stand for any of ``$g$, $v$, $\mu$, $e$", and $\xi_{ab}(1;2)\equiv \langle \delta_a(1)\delta_b(2) \rangle$.
The detailed results are presented in Appendix B. Here, we make a further simplification of the results: we assume the radial separation between the two galaxies is much smaller than their mean comoving radial distance from us $\bar{\chi}$ (i.e. the distant observer approximation), and everything evaluated at $\chi_i$, with $i=1,2$, can be calculated by expanding around $\bar{\chi}$, and keeping only the contributions that are lowest order in $|\chi_i - \bar{\chi}|$. In the following, we give the contributions to $\xi_{\rm obs}(1;2)$ with this simplification.

First, the intrinsic galaxy correlation function is given by
\begin{equation}
\xi_{gg}=\int\frac{d^3k}{(2\pi)^3}e^{i\bf{k}\cdot(\bf{x_1-x_2})}P_{gg}(k,\bar{z}),
\end{equation}
where we suppress the dependence of $(1;2)$ for $\xi_{gg}$, same for the $\xi_{ab}$s given below, $\bf{x_i}$ is the position vector for $(\chi_i,\boldsymbol{\theta_i})$, and $P_{gg}(k,\bar{z})$ is the galaxy power spectrum at $\bar{z}$, with $\bar{z}$ the redshift corresponding to $\bar{\chi}$.

Second, when the effect from peculiar velocities is accounted for, $\xi_{\rm obs}$ has the following corrections
\begin{eqnarray}
\xi_{gv}+\xi_{vg}= 2\bar{f_D} \int\frac{d^3k}{(2\pi)^3}e^{i\bf{k}\cdot(\bf{x_1-x_2})}(\hat{k}\cdot\hat{z})^2 P_{gm}(k,\bar{z}),\\
\xi_{vv}= \bar{f_D}^2 \int\frac{d^3k}{(2\pi)^3}e^{i\bf{k}\cdot(\bf{x_1-x_2})}(\hat{k}\cdot\hat{z})^4 P_{mm}(k,\bar{z}),
\end{eqnarray}
where $\bar{f_D}=f_D(\bar{z})$, with $f_D \equiv d\ln{D}/d\ln{a}$, here $D$ is the linear growth factor of matter perturbation, and $a$ is the scale factor of the universe, $\hat{k}$, $\hat{z}$ are unit vectors pointing respectively in the direction of $\bf{k}$ and to the center of the galaxy sample, and $P_{gm}, P_{mm}$ in turn are the galaxy-matter power spectrum and matter power spectrum. Note we have used the plane-parallel approximation, which assumes the line-of-sight to all the galaxies are the same, parallel to $\hat{z}$. The results given here agree with the literature on the famous Kaiser's effect \cite{Kaiser87}-- the distortion by coherent bulk motions on large scales (in the linear regime), and we neglect the fingers-of-god effect \cite{Jac72}-- the distortion by random motions within collapsed haloes (on small scales). For convenience, the sum of $\xi_{gg}$, $\xi_{gv}+\xi_{vg}$ and $\xi_{vv}$ can be calculated equivalently by using the formulas given in \cite{Ham92} or \cite{MatS96}.

Third, when magnification bias is in addition included, $\xi_{\rm obs}$ has the following extra corrections
\begin{eqnarray}
\xi_{g\mu}+\xi_{\mu g} &=&\frac{3}{2}\Omega_m H_0^2(5\bar{s}-2)(1+\bar{z})|\chi_2-\chi_1|\times\nonumber \\&& \int\frac{d^2k_{\perp}}{(2\pi)^2}e^{i\bf{k_{\perp}}\cdot \bar{\chi}(\boldsymbol{\theta_1}-\boldsymbol{\theta_2})} P_{gm}(k_{\perp},\bar{z}),\label{eqn:xigmu}
\end{eqnarray}
\begin{eqnarray}
\xi_{\mu\mu} &=& \left[\frac{3}{2}\Omega_m H_0^2(5\bar{s}-2)\right]^2\int_0^{\bar{\chi}} d\chi \chi^2 \left(1-\frac{\chi}{\bar{\chi}}\right)^2 \times \nonumber \\&& (1+z)^2 \int\frac{d^2k_{\perp}}{(2\pi)^2}e^{i\bf{k_{\perp}}\cdot \chi(\boldsymbol{\theta_1}-\boldsymbol{\theta_2})} P_{mm}(k_{\perp},z),
\end{eqnarray}
where $\Omega_m$ is the matter density parameter, $H_0$ is the Hubble constant, $\bar{s}=s(\bar{z})$, and $\bf{k_{\perp}}$ is the component of $\bf{k}$ transverse to $\hat{z}$. We have used the Limber approximation, which makes $\xi_{v\mu}$ and $\xi_{\mu v}$ vanish.

Finally, with dust extinction, the following terms should be added to $\xi_{\rm obs}$
\begin{eqnarray}
\xi_{ge}+\xi_{eg}&=&-2.5\bar{s}(1+\bar{z})^{-1}\bar{\rho}_d(\bar{z})f(\bar{z},\lambda_{\rm obs}) \times\nonumber\\&& \int\frac{d^2k_{\perp}}{(2\pi)^2}e^{i\bf{k_{\perp}}\cdot \bar{\chi}(\boldsymbol{\theta_1}-\boldsymbol{\theta_2})} P_{gd}(k_{\perp},\bar{z})\label{eqn:xige},
\end{eqnarray}
where $P_{gd}$ is the galaxy-dust power spectrum. The dependence of $(1+\bar{z})^{-1}$ (the scale factor at $\bar{z}$) comes from converting comoving distance to proper distance in the calculation for $\delta\tau$. As before, we have used the Limber approximation, which also makes $\xi_{ve}$ and $\xi_{ev}$ vanish. We would neglect the corrections from $\xi_{\mu e}$ (or $\xi_{e \mu}$) and $\xi_{ee}$, the rationale being that
extinction is a relatively small effect, and so these corrections are
small compared to the corrections we are keeping, i.e. $\xi_{ee} \ll \xi_{ge}$ and
$\xi_{\mu e} \ll \xi_{\mu g}$.

With our approximation and the symmetry these terms exhibit, their dependence on (1;2) can be simplified, which we summarize as follows: besides $\bar{z}$, $\xi_{gg}$ depends only on the distance between the two points $r=|\bf{x_1-x_2}|$, $\xi_{\mu\mu}$ and $(\xi_{g e}+\xi_{e g})$ depend only on the transverse separation $\delta x_{\perp}=|\bar{\chi}(\boldsymbol{\theta_1}-\boldsymbol{\theta_2})|$, while all others depend on both $\delta x_{\perp}$ and the line-of-sight separation $\delta\chi=|\chi_1-\chi_2|$.
In particular, note how the magnification and extinction distortions
have different shapes: $\xi_{g\mu} + \xi_{\mu g}$ exhibits the
characteristic lensing-induced linear scaling with
the line-of-sight separation $\delta\chi$, while $\xi_{ge} + \xi_{eg}$ does not
depend on it at all.
We assume constant galaxy bias $b_g$ when calculating $P_{gg}$ and $P_{gm}$, i.e. $P_{gg}=b_g^2P_{mm}$ and $P_{gm}=b_gP_{mm}$, and we use the transfer function given by \cite{EH98} and the non-linear prescription given by \cite{smith03} to calculate $P_{mm}$.

\subsection{The Extinction Corrections}

To calculate the extinction corrections, we make use of the recent SDSS observational results by \cite{MenSFR09}, where a positive correlation between the color (g-i) excess of the background quasars and the angular overdensity of the foreground galaxies is found, up to an angular separation of $\simeq 100'$, which suggests the existence of cosmic dust, and by using an extinction curve given by the functional form of \cite{ODon94} with $R_V=3.1$, the one for the standard interstellar dust in the Galactic disk, the result is converted to the following extinction-galaxy cross-correlation
\begin{equation}
\label{AVdeltag}
\langle A_V(\boldsymbol{\theta_1})\delta_g^{2D}(\boldsymbol{\theta_2})
\rangle =2.4\times10^{-3} \left(\frac{|\boldsymbol{\theta_1-\theta_2}|}{1'}\right)^{-0.84},
\end{equation}
where $A_V$ is the V-band extinction from the dust between the observer and the quasars, and $\delta_g^{2D}$ is the angular (2D) overdensity of the galaxies.

With our formulation, the correlation between $A_V$ and $\delta_g^{2D}$ can be calculated by
\begin{eqnarray}
&& \langle A_V(\boldsymbol{\theta_1})\delta_g^{2D}(\boldsymbol{\theta_2}) \rangle \nonumber\\&=&\frac{2.5}{\ln{10}}\int_0^{z_q} dz (1+z)^{-1}\bar{\rho}_d(z)f(z,\lambda_V)\times\nonumber\\&& \frac{\bar{n}(z)}{\bar{n}^{2D}}\int \frac{d^2k_{\perp}}{(2\pi)^2}e^{i\bf{k_{\perp}}\cdot \chi(\boldsymbol{\theta_1}-\boldsymbol{\theta_2})} P_{gd}(k_{\perp},z),
\end{eqnarray}
where $z_q$ is the redshift of the quasars, $\bar{n}(z)$ is the redshift distribution of the galaxies with the normalization of $\bar{n}^{2D}$, their mean surface density,
and we have used the Limber approximation.
Note $A_V$ is defined as $A_V\equiv2.5 \tau_V /{\,\rm ln\,}10$.
In \cite{MenSFR09}, the galaxies' redshift distribution peaks around the mean redshift at $z_*=0.36$, hence we would make the approximation of $\bar{n}(z)/\bar{n}^{2D}\rightarrow \delta(z-z_*)$. Considering $z_q>z_*$, we get
\begin{eqnarray}
&&\langle A_V(\boldsymbol{\theta_1})\delta_g^{2D}(\boldsymbol{\theta_2}) \rangle\nonumber \\&\sim &\frac{2.5}{\ln{10}}(1+z_*)^{-1}\bar{\rho}_d(z_*) f( z_*,\lambda_V)\times\nonumber\\&& \int \frac{d^2k_{\perp}}{(2\pi)^2}e^{i\bf{k_{\perp}}\cdot \chi_* (\boldsymbol{\theta_1}-\boldsymbol{\theta_2})} P_{gd}(k_{\perp}, z_*),\label{eqn:AVobs}
\end{eqnarray}
where $\chi_*$ is the comoving radial distance corresponding to $z_*$.

Comparing Eqn~(\ref{eqn:AVobs}) with Eqn~(\ref{eqn:xige}), we find when $\bar{z}=z_*$, and when $\lambda_{\rm obs}=\lambda_V$,
\begin{equation}
\xi_{ge}+\xi_{eg}=-\ln{10} \bar{s} \langle A_V(\boldsymbol{\theta_1})\delta_g^{2D}(\boldsymbol{\theta_2}) \rangle \, .\label{eqn:xigeV}
\end{equation}
Thus, the result of \cite{MenSFR09} can be directly translated into
the quantity we are interested in. Of course, this direct translation works
only when the mean properties of the galaxies (e.g. redshift, clustering bias
and so on) coincide with those used in \cite{MenSFR09}. We thus need to
extrapolate. First, we extrapolate in angle. Keeping everything (e.g. redshift
and so on) fixed, the result summarized in Eqn (\ref{AVdeltag}) is applicable
for angular separations up to $100'$. Beyond that, we assume
the shape of the two-point function follows that of matter, in other words
that
\begin{equation}
[\xi_{ge}+\xi_{eg}] (z_*)\propto \int \frac{d^2k_{\perp}}{(2\pi)^2}e^{i\bf{k_{\perp}}\cdot \chi_*(\boldsymbol{\theta_1}-\boldsymbol{\theta_2})} P_{mm}(k_{\perp},z_*) \, ,
\end{equation}
with a normalization that matches the observed result at $100'$.
Next we extrapolate in bias. Since the galaxy sample used in \cite{MenSFR09}
has a clustering bias $\simeq 1$, we multiply Eqn (\ref{eqn:xigeV}) by
$\bar b_g$ to obtain $\xi_{ge} + \xi_{eg}$ for
a different galaxy sample with a mean bias of $\bar b_g$.
Lastly, we extrapolate in redshift. Based on Eqn~(\ref{eqn:xige}), we assume the following redshift scaling
for a fixed comoving transverse separation $\delta x_\perp$:
\begin{equation}
[\xi_{ge}+\xi_{eg}] (\bar z) \propto(1+\bar{z})^{-1}\bar{\rho}_d(\bar{z})D(\bar{z})^2.\label{eqn:xigeevol}
\end{equation}
The number count slope $\bar s$ and the clustering bias $\bar b_g$
should be redshift dependent as well -- we will systematically vary these
two parameters in the following computations to illustrate the range of possibilities.
It should be stressed that
the scaling should also take into account of the redshift dependence of the clustering bias of dust and the extinction efficiency $f$ for a given $\lambda_{\rm obs}$ (in our case, $\lambda_{\rm obs}=\lambda_V$). In this paper, without detailed modeling of clustering and extinction properties of the dust, we would simply neglect the $z$-dependence of both quantities, (we notice that, for $f$, this is supported by the slowly-varying extinction curve in the visible range, if the dust is like that in the Galactic disk, see e.g. \cite{ODon94}.) and hope our procedure of systematically exploring variations in $\bar s$ and $\bar b_g$ is sufficient to bracket the range of possibilities.

For the redshift dependence of $\bar{\rho}_d$, we follow \cite{Cor06}, and assume the dust particles are ejected to the intergalactic medium with a constant yield when new stars are born, so
\begin{equation}
\bar{\rho}_d(z)\propto (1+z)^3\int_z^{z_s}\frac{\dot{\rho}_{\rm SFR}(z')dz'}{(1+z')H(z')},
\end{equation}
where we set the star formation beginning at redshift $z_s=10$ \cite{Cor06}, and for the cosmic star formation rate (SFR) $\dot{\rho}_{\rm SFR}$, the mass of baryons that form into star per unit comoving volume per unit proper time, we use the results from \cite{MadP00} after converting to our cosmology according to \cite{PorM01}.

\section{The Anisotropic Distortion}
\label{sec:anisotropy}

In this section, we show our results on the extinction distortion of the galaxy correlation function, and compare it with other anisotropic distortions. We first list the values of the parameters used in our calculation. Then by plotting contours of the galaxy correlation function, we explicitly display the anisotropic distortions. The results are then extensively studied in the paragraphs following the italicized titles.

The corrections from dust extinction depend on the properties of galaxy sample through three parameters, $\lambda_{\rm obs}$, $s$ and $b_g$. We have set $\lambda_{\rm obs}=\lambda_V$ for the calculation in this paper, and for the purpose of illustration, we set $s=1.5$ and $b_g=2$, similar to those for the SDSS Luminous Red Galaxies (LRGs) \cite{GCH09}. We discuss the dependence of the results on these parameters at the end of this section. For the cosmological model, we choose it to be the best-fit flat $\Lambda$CDM model from the WMAP 7-year results that has: matter density $\Omega_m=0.27$, baryon density $\Omega_b=0.045$, Hubble constant $h=0.71$, normalization of the power spectrum $\sigma_8=0.8$, and the spectral index $n_s=0.96$ \cite{WMAP7}.

\begin{figure*}[htb]
\vspace{-25mm}
\resizebox{160mm}{!}{\includegraphics{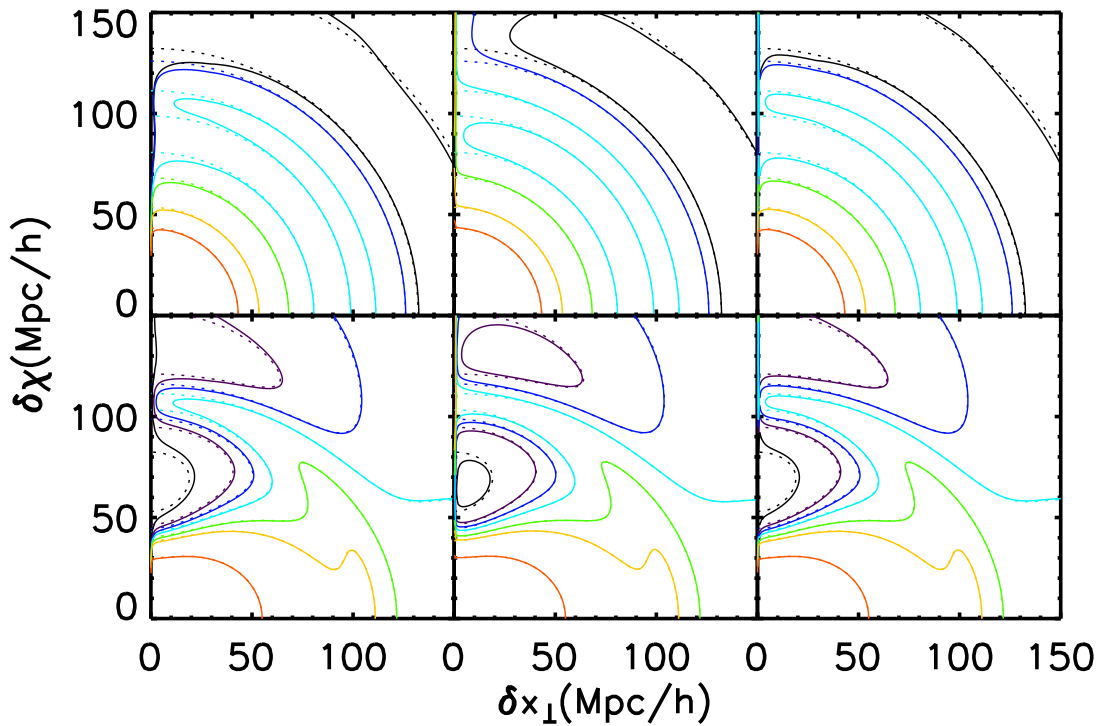}}
\vspace{-2mm}
\caption{\label{fig:contz036} Contours of the galaxy correlation function divided by $b_g^2$ at $\bar{z}=0.36$. $\delta\chi$ and $\delta x_{\perp}$ are the line-of-sight and transverse separations respectively. Dotted lines in the upper three panels are for the intrinsic galaxy correlation function, while those in the lower three panels include redshift distortion as well. In each row, from left to right, the solid lines are the same with the dotted lines except that dust extinction, magnification bias, and both dust extinction and magnification bias are respectively added in. The colors from black through the rainbow to red represent contour levels at ($-1.5\times 10^{-4}, 0, 0.001, 0.002, 0.005, 0.01$) in the upper three panels, and ($-0.002, -0.001, -0.0005, 0, 0.001, 0.002, 0.01$) in lower three panels.}
\end{figure*}

\begin{figure*}[htb]
\vspace{-25mm}
\resizebox{160mm}{!}{\includegraphics{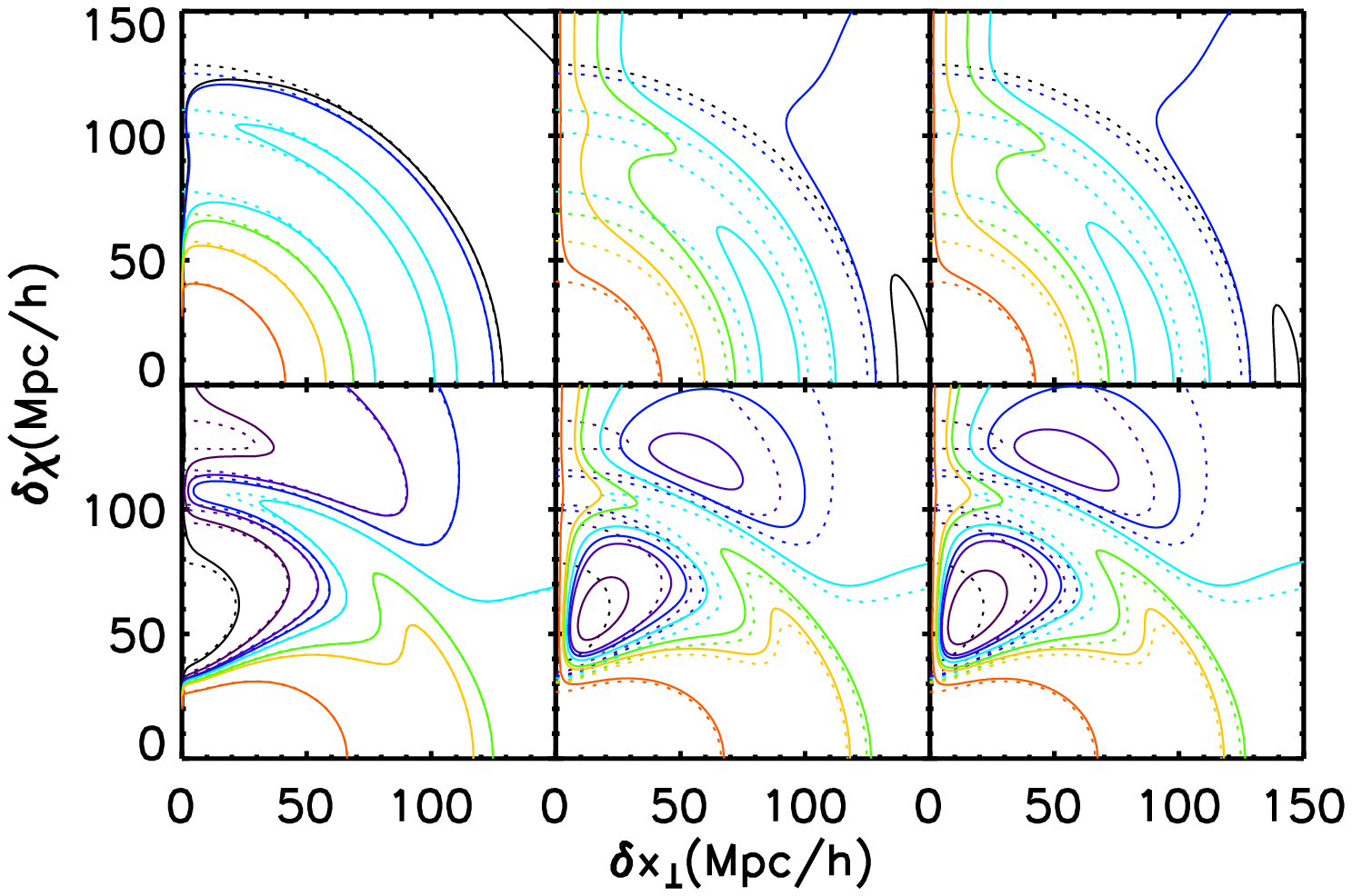}}
\vspace{-2mm}
\caption{\label{fig:contz2} Same as Figure~\ref{fig:contz036} but for $\bar{z}=2$. The colors from black through the rainbow to red represent contour levels at ($-3\times 10^{-5}, 0, 0.0003, 0.0005, 0.001, 0.003$) in the upper three panels, and ($-0.001, -0.0005, -0.00025, -0.00015, 0, 0.0003, 0.0005, 0.002$) in lower three panels.}
\end{figure*}

In Figure~\ref{fig:contz036}, we show the contours of galaxy correlation function after dividing by $b_g^2$ at $\bar{z}=0.36$. The dotted lines in the upper three panels are for the intrinsic galaxy correlation function $\xi_{gg}$, while those in the lower three panels include redshift distortion as well. In each row, from left to right, the solid lines are the results with dust extinction, magnification bias, and both dust extinction and magnification bias added in, compared to the dotted lines. So by comparing the solid lines with dotted lines in the left two panels, we can see distortions caused by dust extinction to the intrinsic galaxy correlation function with redshift distortion included (lower panel) or not (upper panel), similar for the middle and the right two panels, where the distortions caused by magnification bias and the combined effect of magnification bias and dust extinction can be seen.

{\it{The Extinction Anisotropy.}} The upper left panel explicitly shows that dust extinction introduces anisotropy to the galaxy correlation function in the ($\delta x_{\perp}, \delta\chi$) plane, as is expected from our calculation: Eqn~(\ref{eqn:xige}) depends only on $\delta x_{\perp}$, as a result of Limber approximation. For a given $\delta x_{\perp}$, since the amplitude of the intrinsic correlation generally decreases when $\delta\chi$ (or $r$) increases, while the extinction corrections remain the same, this leads to a bigger effect of dust extinction at a larger $\delta\chi$ (or $r$). Similarly, for a given $r$, since the amplitude of the extinction corrections generally increases when $\delta x_{\perp}$ decreases, while the intrinsic correlation remains the same, a larger effect of dust extinction is expected at smaller $\delta x_{\perp}$, or when the alignment of the separation is closer to the line of sight (LOS). Therefore, the distortion by dust extinction is expected to be most significant at large separations along the LOS, which agrees with our findings in the upper left panel of Figure~\ref{fig:contz036}.

{\it{Comparison with Magnification Bias.}} The upper middle panel of Figure~\ref{fig:contz036} allows us to make a comparison of the extinction anisotropy with that from the magnification bias, which has been well studied by earlier works, e.g. \cite{Mat00,HuiGL07a}. Same as dust extinction, the effect of magnification bias is also more important when the (LOS) separation is larger and the alignment of the separation is closer to the LOS. In a similar way as before, this can be understood by the following: for a given $\delta x_{\perp}$, in general, the amplitude of the total corrections from magnification bias increases when $\delta\chi$ increases, ($\xi_{\mu\mu}$ remains the same, $\xi_{g\mu}+\xi_{\mu g}$ increases linearly with $\delta\chi$,) while that for the intrinsic correlation decreases, leading to a bigger effect of magnification bias at a larger $\delta\chi$; for a given $r$, the amplitude of the total corrections generally increases when $\delta x_{\perp}$ decreases, while the intrinsic correlation remains the same, leading to a bigger effect at a smaller $\delta x_{\perp}$, see also \cite{Mat00,HuiGL07a}.
However, different from dust extinction, the anisotropy from magnification bias has opposite features---the contours are distorted to the opposite sides of their intrinsic locations.

An examination of Eqn~(\ref{eqn:xigmu}) and Eqn~(\ref{eqn:xige}) tells us that $(\xi_{g\mu}+\xi_{\mu g})$ and $(\xi_{ge}+\xi_{eg})$ have similar expressions: both are given as a 2D integral of the matter power spectrum (note we assume constant bias for both the galaxies and the dust) multiplied by a factor. This is the result of the fact that both $\kappa$ and $\delta\tau$ are weighted integrals of fluctuations along the LOS, and that we adopt Limber approximation in the calculation.
There are two differences between the two:
first, magnification bias has the characteristic linear dependence on the LOS separation $\delta\chi$ from lensing; second dust extinction has a dependence on the
observed wavelength. These differences can be exploited to separate out the
two effects from data.

For magnification bias, the sign of the overall multiplicative factor is determined by $(5s-2)$, while for dust extinction, it is always negative. With our choice of $s=1.5$, $(5s-2)=5.5>0$, so $(\xi_{g\mu}+\xi_{\mu g})$ and $(\xi_{ge}+\xi_{eg})$ always have opposite signs in our calculation. Since the correction from $\xi_{\mu\mu}$ is less important at a lower redshift compared with those from $(\xi_{g\mu}+\xi_{\mu g})$ \cite{HuiGL07a}, the analysis here explains the opposite anisotropic features from magnification bias and dust extinction throughout the displayed regions in Figure~\ref{fig:contz036}. Specifically, with our choice of cosmology, the 2D integral of matter power spectrum is positive when $\delta x_{\perp}\lsim 115 h^{-1}$Mpc, and negative otherwise. It is the same for the sign of $(\xi_{g\mu}+\xi_{\mu g})$, while the opposite holds true for that of $(\xi_{ge}+\xi_{eg})$. Therefore, the contours are distorted to where $\xi_{gg}$ has larger (smaller) values by dust extinction (magnification bias) when $\delta x_{\perp}\lsim 115 h^{-1}$Mpc, and to where $\xi_{gg}$ has smaller (larger) values when $\delta x_{\perp}\gsim 115 h^{-1}$Mpc.

{\it{Combination with Magnification Bias.}} Due to the canceling effect between dust extinction and magnification bias in our calculation, we can see from the upper right panel that the anisotropy from the combination of these two is weakened to some extent, with the final features dominated by those from dust extinction.

{\it{Including Redshift Distortion.}} Since in redshift space, the most significant anisotropic features in the galaxy correlation function are caused by redshift distortion, we replot the upper three panels of Figure~\ref{fig:contz036} in the lower three ones by including redshift distortion in all the contours, for a more realistic view of the anisotropies we saw before. According to the prediction of the Kaiser's effect, which is represented by the dotted lines in the lower three panels, the galaxy correlation function has a quadrupole and a hexadecapole component of anisotropy \cite{Ham92, MatS96}, with the magnitudes controlled by the linear redshift distortion parameter $\beta\equiv f/b_g$.
Comparing the dotted lines with the solid lines, we can see that, same as before, dust extinction distorts the contours to where they had larger values for most of the displayed regions, while magnification bias does the opposite, so the distortions from their combination are reduced, with the final results still dominated by dust extinction. Along each contour, dust extinction would be more important where $\delta x_{\perp}$ is smaller, while magnification bias would be more important where both $\delta\chi$ is larger and $\delta x_{\perp}$ is smaller.

{\it{Dependence on Redshift.}} To see the redshift dependence of the anisotropic features, we show the above contours for a higher redshift $\bar{z}=2$ in Figure~\ref{fig:contz2}. For dust extinction, its relative importance to the intrinsic correlation scales with redshift approximately by a factor of $(1+\bar{z})^{-1}\bar{\rho}_d(\bar{z})$, see Eqn~(\ref{eqn:xigeevol}), which, according to our model for the evolution of dust, increases with redshift until $\bar{z}\simeq 1.2$, where it has a $\sim 60\%$ increase from its value at $\bar{z}=0.36$, then decreases gradually. At $\bar{z}=2$, the factor is $\sim 30\%$ larger than at $\bar{z}=0.36$, so the extinction anisotropy shown in the upper left panel of Figure~\ref{fig:contz2} gets slightly stronger.

At the same time, the relative importance of magnification bias to the intrinsic correlation also becomes larger, as can be directly seen from the much stronger anisotropy shown in the upper middle panel. Both the corrections from $(\xi_{g\mu}+\xi_{\mu g})$ and that from $\xi_{\mu\mu}$ become more important at a higher redshift: for the former, its ratio to $\xi_{gg}$ scales with $\bar{z}$ roughly by $(1+\bar{z})$, hence increases with $\bar{z}$; for the latter, the ratio also increases, for $\xi_{\mu\mu}$ increases as a sum along a longer LOS, while $\xi_{gg}$ decreases. The latter also becomes more important compared to the former, see \cite{HuiGL07a}, so, different from at the low redshift, the total corrections for $\delta x_{\perp}\gsim 115 h^{-1}$Mpc at $\bar{z}=2$ is now also positive, and all the displayed contours are distorted to where $\xi_{gg}$ has smaller values. At this high redshift, the combination of dust extinction and magnification bias is absolutely dominated by magnification bias, with its effect mildly weakened by dust extinction when $\delta x_{\perp}\lsim 115 h^{-1}$Mpc, while strengthened when $\delta x_{\perp}\gsim 115 h^{-1}$Mpc. Similar results hold when redshift distortion is included.

{\it{Dependence on Galaxy Sample.}} Finally, we discuss the dependence of the anisotropic features on the properties of the galaxy sample. For dust extinction, the anisotropy depends on the ratio of $s$ to $b_{g}$, and is stronger for galaxy samples with larger $s$ or smaller $b_g$. For magnification bias, the anisotropy is controlled by the factor of $(5s-2)/b_g$, which vanishes for $s=0.4$ where $(5s-2)$ switches sign. In our calculation, had we chosen $s<0.4$, $(\xi_{g\mu}+\xi_{\mu g})$ would have the same sign as $(\xi_{ge}+\xi_{e g})$, and the combination of magnification bias and dust extinction would instead add to each other at the low redshift. The anisotropy from redshift distortion is known to be controlled by $\beta$ that depends on the sample through $1/b_g$. So for galaxy samples with different $s$ and $b_g$, the relative importance of these anisotropic effects would be different. Besides $s$ and $b_g$, the anisotropy from dust extinction in addition depends on the bandpass used to observe the sample. The shorter wavelength the bandpass allows, the more extinction the dust particles cause, and the stronger the anisotropy would be. Altogether, the different dependence on the sample parameters of these anisotropic effects provide an opportunity to potentially isolate them from one another.

\section{Effects on Cosmological Probes}
\label{sec:systematic}

In this section, we study the effect of dust extinction on cosmological probes through measurements of the galaxy correlation function. We consider two of these: one is the BAO peak, which serves as a standard ruler and probes the geometry of the universe, the other is the linear redshift distortion parameter $\beta$, which directly probes the growth rate of the universe's structure.

\subsection{The BAO Peak}

\begin{figure*}[htb]
\centering
\subfigure[]
{\includegraphics[scale=0.5]{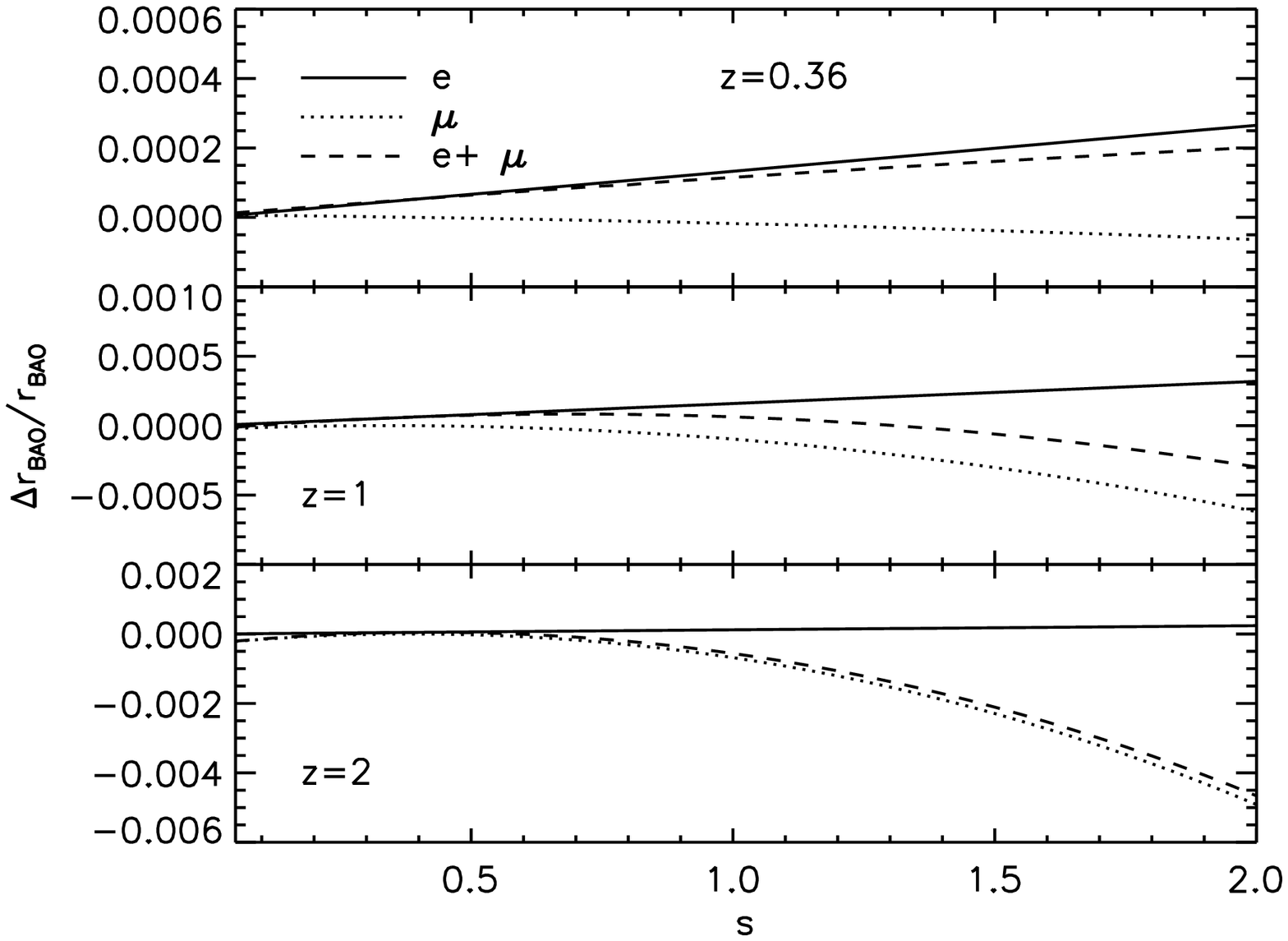}\label{subfig:baovs}}
\hspace{5mm}
\subfigure[]
{\includegraphics[scale=0.5]{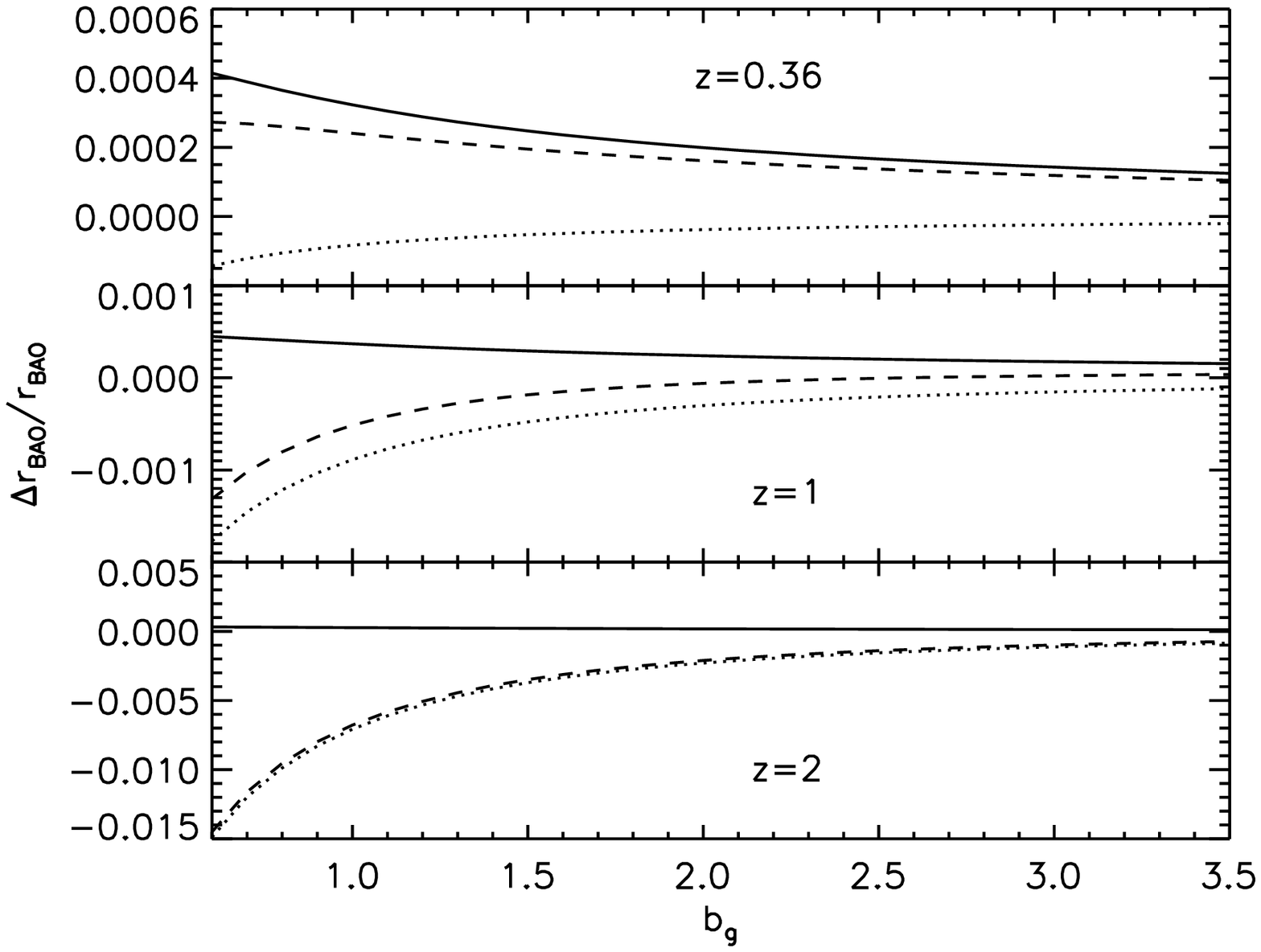}\label{subfig:baovb}}
\caption{\label{fig:bao} Fractional shift of the BAO peak from the monopole of the galaxy correlation function by dust extinction (the solid lines, labeled by ``$e$''), magnification bias (the dotted lines, labeled by ``$\mu$'') and their combined effects (the dashed lines, labeled by ``$e+\mu$''). In Figure~\ref{subfig:baovs}, we keep $b_g$ at our default choice of $b_g=2$ and vary $s$, while in Figure~\ref{subfig:baovb}, we set $s$ at the default value of 1.5 and vary $b_g$. From top to bottom, the three panels in both figures are for $z=0.36$, $z=1$ and $z=2$ respectively. Note redshift distortion is included in the calculation for all the monopoles, which by itself does not shift the monopole BAO peak according to the prediction of the Kaiser's effect \cite{Kaiser87,Ham92,MatS96}.}
\end{figure*}

The BAO peak has been detected from the monopole of the galaxy correlation function, i.e. the average over the alignment of the separation vector, see e.g. \cite{Eis05,Mar08,CabG09,Kaz10a}. In Figure~\ref{fig:bao}, by identifying the BAO peak as a local maximum in the monopole of $\xi_{\rm obs}$, we show the fractional shift of the peak location $\Delta r_{\rm BAO}/r_{\rm BAO}$ caused by dust extinction (the solid lines), and for comparison, by magnification bias (the dotted lines) and the combination of the two (the dashed lines). The results are given for three different redshifts: $z=0.36$ (upper panels), $z=1$ (middle panels) and $z=2$ (lower panels). In Figure~\ref{subfig:baovs}, we vary $s$ and keep $b_g=2$, while in Figure~\ref{subfig:baovb}, we vary $b_g$ and keep $s=1.5$. Note redshift distortion is included when we calculate all the monopoles, which by itself does not shift the monopole BAO peak according to the prediction of the Kaiser's effect \cite{Kaiser87,Ham92,MatS96}.

The scale-dependent correction to the monopole from dust extinction shifts $r_{\rm BAO}$ to larger values. This is understandable, since, as a negative but increasing component, the correction would shift the local maximum to the right.
The shift is larger when $s$ is larger or $b_g$ is smaller, consistent with our analysis above for the extinction distortion. As a positive but decreasing component, the correction from magnification bias shifts $r_{\rm BAO}$ to smaller values, except at low redshift and when $s<0.4$, and the shift is larger when $|5s-2|$ is larger or $b_g$ is smaller.
Dust extinction tends to act in the opposite direction,
so the combination of dust extinction and magnification bias helps to reduce the shift in $r_{\rm BAO}$, except at low redshift and when $s<0.4$. For most of the cases, the much stronger redshift evolution of the effect from magnification bias, compared to that for dust extinction, makes the latter negligible at high redshift, though it is the dominating effect at low redshift.
However, the fractional shift of $r_{\rm BAO}$ by dust extinction is on the order of $10^{-4}$, which makes it unlikely to be
an important factor for probes of the monopole BAO peak
even at low redshift.

There has been a claimed detection of the BAO peak from
the LOS galaxy correlation function by
\cite{GCH09}, which was disputed by \cite{Kaz10b}.
The issue is subtle: \cite{CG11} showed
using a large set of simulations
that even for the {\it monopole} correlation function,
the BAO peak detection is only marginal; yet, useful
cosmological constraints can be inferred once combined
with external data. Indeed, the LOS measurements
of \cite{GCH09} and \cite{Kaz10b} are consistent with
each other. It is the interpretation (of how the data should
be used) that differs.
For our purpose in this paper, it suffices to note that
the dust extinction correction has no LOS dependence
and so would not shift the LOS BAO peak at all.

\subsection{Redshift Distortion}

\begin{figure}[htb]
\resizebox{90mm}{!}{\includegraphics{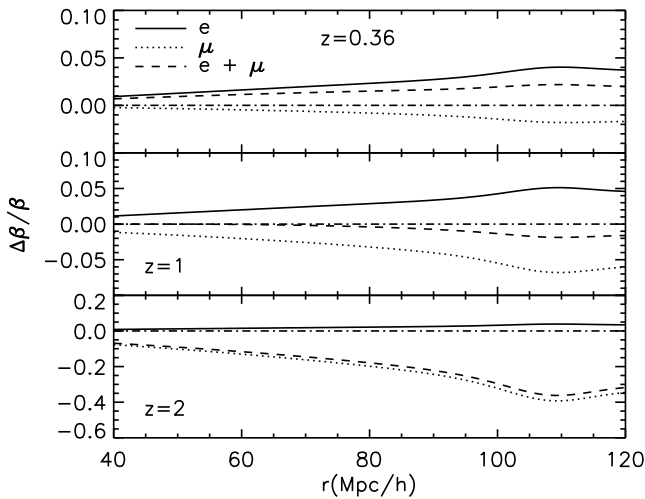}}
\caption{\label{fig:beta} Fractional changes in the linear redshift distortion parameter $\beta$, inferred through Eqn~(\ref{eqn:getbeta}), as a function of the distance separation $r$. The solid lines show the changes caused by dust extinction (labeled by ``$e$''), the dotted lines are those by magnification bias (labeled by ``$\mu$''), and the dashed lines are by their combination (labeled by ``$e+\mu$''). To see the changes more easily, we also show $\Delta \beta=0$ using dot-dashed lines. From top to bottom, the three panels are for $z=0.36$, $z=1$ and $z=2$ respectively. The values for $s$ and $b_g$ are set at our default choice.}
\end{figure}

According to the prediction of the Kaiser's effect \cite{Kaiser87,Ham92,MatS96}, the redshift distortion parameter $\beta$ can be inferred from the anisotropy of the galaxy correlation function through \cite{Ham92}
\begin{equation}
\frac{\xi_2(r)}{\xi_0(r)-\bar{\xi}_0(r)}= \frac{\frac{4}{3}\beta+\frac{4}{7}\beta^2}{1+\frac{2}{3}\beta+\frac{1}{5}\beta^2},\label{eqn:getbeta}
\end{equation}
where $\xi_0$ and $\xi_2$ are the monopole and quadrupole of $\xi_{\rm obs}$, and $\bar{\xi}_0$ is the volume average of $\xi_0$, given by
\begin{equation}
\bar{\xi}_0(r)=\frac{3}{r^3}\int_0^{r}\xi_0(r')r'^2dr'.
\end{equation}
However, when the effects of dust extinction and magnification bias are taken into account, their corrections to the monopole and quadrupole would shift the inferred $\beta$ from its true value, and moreover, the scale-dependence of the corrections may cause the inferred $\beta$ to be scale-dependent too. This is what we find in Figure~\ref{fig:beta}, where the fractional changes in $\beta$ caused by dust extinction (the solid lines), magnification bias (the dotted lines) and their combination (the dashed lines) are presented at $z=0.36$ (the upper panel), $z=1$ (the middle panel), and $z=2$ (the lower panel) respectively. The results are obtained with our default values: $s=1.5$ and $b_g = 2$.

As can be seen, the changes by dust extinction and magnification bias are in opposite directions, with $\beta$ getting bigger for the former and smaller for the latter, so the changes by their combination tend to be reduced. If we define $Q\equiv \xi_2/(\xi_0-\bar{\xi}_0)$, we find dust extinction introduces $\Delta(\xi_0-\bar{\xi}_0)>0$ ($\Delta \xi_0$ increases with $r$), $\Delta \xi_2<0$ ($\Delta\xi$ increases when the alignment of the separation is closer to the transverse), while with the prediction of the Kaiser's effect, $(\xi_0-\bar{\xi}_0)$ and $\xi_2$ are both negative (same as $(\xi_{gg}-\bar{\xi}_{gg})$, see e.g. \cite{Ham92}), so dust extinction leads to $\Delta Q>0$. When we infer $\beta$ from $Q$ through Eqn~(\ref{eqn:getbeta}), we have $dQ/d\beta>0$, so this explains why dust extinction causes $\beta$ to be larger. With $s > 0.4$, magnification bias introduces opposite changes, so the smaller value of the inferred $\beta$ when magnification bias is included can be understood in a similar way.

Our calculation also shows that the fractional changes by dust extinction are on the order of a few percent, and vary mildly with redshift, while those by magnification bias are on the level of $\sim 1\%$ at $z=0.36$, but can grow up to $\sim 40\%$ at $z=2$, which turns it from a sub-dominant effect at low redshift to a dominating effect at high redshift. These fractional changes indicate that both dust extinction and magnification bias would be non-negligible for redshift distortion probes for the purpose of precision cosmology, e.g. \cite{wigglez,boss,bigboss,euclid}, especially at low redshift for the former and at high redshift for the latter, though the canceling effect between these two helps to reduce these systematics.

Finally, we point out that the scale dependence of the inferred $\beta$ may be mistakenly interpreted as an indication for modified gravity, which, different from General Relativity (GR), can give a scale-dependent growth factor, and lead to wrong conclusions for tests of GR through probes of the redshift distortion parameter \cite{Acq08,Acq10,ZhaL07,ZhaB08}, if these systematics are neglected.


\section{Discussion}
\label{sec:discuss}

Inhomogeneities in the extinction of the galaxies' fluxes by cosmic dust, whose existence is recently detected by \cite{MenSFR09}, modify the distribution of a flux-selected galaxy sample. This is similar to what the inhomogeneous magnification of the galaxies' fluxes by gravitational lensing does to the galaxies' distribution, but with opposite effect on average.
Hence, in addition to the Alcock-Paczynski effect, redshift distortion,  magnification bias, dust extinction is a fourth effect that would create an anisotropic distortion to the galaxy correlation function for a flux-limited selection. In this paper, we have studied this extinction distortion to the galaxy correlation function, and evaluated its effect on cosmological probes such as the BAO and the linear redshift distortion.

We use the extinction-galaxy cross correlation found by \cite{MenSFR09} to calculate the corrections to the galaxy correlation function from dust extinction. We extrapolate their results to larger scales by assuming dust traces galaxies, and to other redshifts by assuming the evolution of dust follows that of the stars. With the choice of $\lambda_{\rm obs}=\lambda_V$, $s=1.5$, $b_g=2$, we show the anisotropic extinction distortion together with that from magnification bias and redshift distortion in Figure~\ref{fig:contz036} and Figure~\ref{fig:contz2}.

We find the distortion by dust extinction alone is most significant along the LOS and at large separations, which is similar to that by magnification bias. Their precise shapes are different
though. Lensing induces a correction to the correlation
function that rises with the LOS separation, while extinction does not.
The correction from dust extinction depends only on the transverse separation $\delta x_{\perp}$, and with our choice of cosmology, it is negative when $\delta x_{\perp}\lsim 115 h^{-1}$Mpc, positive otherwise. With our choice of $s>0.4$, the correction is almost always opposite in sign to that from magnification bias, leading to the opposite anisotropic features seen in the distortions by these two effects. So, the distortion by their combined effect tends to be reduced. The extinction distortion evolves with redshift approximately by $(1+\bar{z})^{-1}\bar{\rho}_d(\bar{z})$, which is much milder than the evolution of the lensing distortion.
At low redshifts ($\bar z \lsim 1$),
the extinction distortion tends to be more important
than lensing, while the opposite is true at high redshifts.

By identifying the BAO peak as a local maximum, we find the scale-dependent correction from dust extinction to the monopole of the galaxy correlation function shifts the monopole BAO peak to larger scales, but the shift is on the order of $10^{-4}$, and it does not change much when varying $s$, $b_g$ and $\bar{z}$. At the same time, the scale-independent correction from dust extinction to the LOS correlation function evaluated at a fixed $\delta x_{\perp}$ does not shift the LOS BAO peak at all. So for probes of the BAO, dust extinction is probably a negligible effect.

The anisotropic extinction distortion also introduces biases in the linear redshift distortion parameter $\beta$, inferred from the monopole and quadrupole of the observed galaxy correlation function according to the prediction of the Kaiser's effect. We find with dust extinction, the inferred $\beta$ is bigger than the true value by up to a few percent, and the shift varies mildly with redshift, while with magnification bias ($s>0.4$), $\beta$ is smaller than the true value by up to percent level at low redshift ($z \lsim 1$), but to $\sim 40\%$ at high redshift. This suggests both effects are non-negligible for precision probes of $\beta$, especially for extinction at low redshift and lensing at high redshift, though their combination tends to reduce the overall
shift in $\beta$.
With these two effects, the inferred $\beta$ also becomes scale-dependent, which should be taken into account for tests of GR through the scale-dependence of $\beta$ (the growth factor). Our analysis on $\beta$ can be extended to Fourier space. 
It is possible that the changes in $\beta$ in Fourier space would be smaller, as suggested by earlier works of \cite{HuiGL07b}, who find that, for magnification bias, its impact on galaxy correlation is less severe in Fourier space than in real space. However, we would leave a rigorous study for future work.

The extinction distortion (extinction correction normalized by
the intrinsic galaxy correlation) scales with the properties of the galaxy sample as $s/b_g$, so it is more significant for galaxy samples with larger $s$ or smaller $b_g$, and it also depends on $\lambda_{\rm obs}$, the bandpass used to observe the sample, shorter $\lambda_{\rm obs}$, stronger distortion. This is different from the distortion by magnification bias, which depends on $(5s-2)/b_g$, and has no dependence on $\lambda_{\rm obs}$.
Moreover, these two distortions have different shapes, with
lensing exhibiting the signature linear dependence on the LOS
separation, while extinction depends exclusively on the transverse
separation.
These differences can be used to separate the two effects, to allow a simultaneous study of both the cosmic extinction and the cosmic magnification \cite{ZhaP05}, which we hope to explore in future work.

\begin{acknowledgments}

We thank J$\ddot{\rm o}$rg Dietrich and Guilin Liu for helpful discussions, and Zolt$\acute{\rm a}$n Haiman, Dragan Huterer for useful comments on the manuscript. W.F. is supported by the NSF under contract AST-0807564, and by the NASA under contract NNX09AC89G. L.H. is supported by the DOE
DE-FG02-92-ER40699 and the NASA NNX10AH14G.

\end{acknowledgments}

\vfill
\bibliographystyle{physrev}
\bibliography{draft}

\onecolumngrid
\appendix
\begin{center}
  {\bf APPENDIX A}
\end{center}

In this Appendix, we give our derivation for the observed galaxy overdensity in the presence of peculiar velocities, gravitational lensing and dust extinction, as given by Eqn~(\ref{eqn:deltaobs})-(\ref{eqn:deltae}) in \S \ref{subsec:xiobs}.

Suppose we are making observations in a smooth universe without dust, and galaxies can be thought of as test particles that are at rest on the comoving grid, the number of galaxies we observe whose positions are within the comoving volume element around ($\chi_g, \boldsymbol{\theta_g}$) with a radial extent $d\chi_g$ and an angular extent $d^2\theta_g$, and whose fluxes are within the range of [$f_g^{\lambda_g}$, $f_g^{\lambda_g}+df_g^{\lambda_g}$] is $\Phi_g(\chi_g, \boldsymbol{\theta_g};f_g^{\lambda_g})\chi_g^2 d\chi_g d^2\theta_g df_g^{\lambda_g}$. Here, the observed flux $f_g$ is labeled by the source rest-frame wavelength $\lambda_g$, while its dependence on the observer rest-frame wavelength $\lambda_{\rm obs}$ is omitted, because in the following, we are about to compare this hypothetical observation with the real observation, while $\lambda_{\rm obs}$ is the same for the both (same filter), $\lambda_g$ is not. Note, we use the subscript ``g" to label the quantities from our hypothetical observation, and throughout this paper, we assume a flat universe.

In the real universe with perturbations and dust, the observed coordinates and fluxes of these galaxies will be shifted according to the following
\begin{eqnarray}
\chi_g &\rightarrow& \chi=\chi_g+\frac{(1+z_g)v_{\parallel}}{H(z_g)}\label{eqn:chiobs},\\
\boldsymbol{\theta_g} &\rightarrow& \boldsymbol{\theta}=\boldsymbol{\theta_g}+ \delta\boldsymbol{\theta_g},\\
f_g^{\lambda_g} &\rightarrow& f^{\lambda}=e^{-\tau} A f_g^{\lambda_g}\label{eqn:fobs},
\end{eqnarray}
where the change in $\chi_g$ follows from that in the redshift $z_g$ ($\equiv \lambda_{\rm obs}/\lambda_g-1$), which is a direct observable and shifts as
\begin{equation}
z_g \rightarrow z = z_g+(1+z_g)v_{\parallel},
\end{equation}
where we have set the speed of light $c=1$, kept only to the first order of perturbations, as we do throughout the paper, and we have included only the change caused by the line-of-sight peculiar velocity of the galaxies $v_{\parallel}$; the apparent angular position is displaced by gravitational lensing by an amount of $\delta\boldsymbol{\theta_g}$; the observed flux is changed in two ways: first, it is magnified by gravitational lensing by a factor of $A \equiv \det[{\partial \theta_{g}^i}/{\partial \theta^j}]^{-1}$, and second, it is reduced by dust extinction by a factor of $e^{-\tau}$, with $\tau$ the extinction optical depth.
For additional effects that affect the galaxies' flux, see
\cite{Bon06,HG06,Bon08} .
We have suppressed in the above equations the position dependence of $(\chi_g, \boldsymbol{\theta_g})$ for $v_{\parallel}$, $\delta\boldsymbol{\theta_g}$, $\tau$, and $A$,  and also the wavelength dependence of $\lambda_{\rm obs}$ for $\tau$.
Despite the changes of these observables, the number of the galaxies should be conserved, so we have
\begin{equation}
\Phi(\chi, \boldsymbol{\theta};f^{\lambda})\chi^2 d\chi d^2\theta df^{\lambda}=\Phi_g(\chi_g, \boldsymbol{\theta_g};f_g^{\lambda_g})\chi_g^2 d\chi_g d^2\theta_g df_g^{\lambda_g}.\label{eqn:ngconserve}
\end{equation}

Given the results from a real observation, a galaxy sample can be selected by specifying the selection efficiency function $\epsilon(f^{\lambda})$. As an example, for a selection with a sharp faint-end cutoff at the limiting flux of $f^{\lambda}_{\rm min}$, $\epsilon(f^{\lambda})$ is just the heaviside step function $\Theta(f^{\lambda}-f^{\lambda}_{\rm min})$. The observed galaxy density of the sample is then given by
\begin{equation}
n(\chi,\boldsymbol{\theta})=\int \epsilon(f^{\lambda})\Phi(\chi,\boldsymbol{\theta};f^{\lambda})d f^{\lambda}.
\end{equation}
Note, hereafter, we would call only the quantities from the real observation the ``observed" ones, while those from our hypothetical observation (in a smooth universe) would be called the ``intrinsic" ones. By using Eqns~(\ref{eqn:chiobs})-(\ref{eqn:fobs}) and Eqn~(\ref{eqn:ngconserve}), we find
\begin{equation}
n(\chi,\boldsymbol{\theta})=(1-2\kappa)\left[1-\frac{(1+z_g)}{H(z_g)} \frac{\partial v_{\parallel}}{\partial \chi_g}\right]\int \epsilon(e^{-\tau}A f_g^{\lambda_g})\Phi_g(\chi_g,\boldsymbol{\theta_g};f_g^{\lambda_g})d f_g^{\lambda_g},
\end{equation}
where the first bracket on the right hand side of the equation comes from $d^2\theta_g/d^2\theta=1/A$, and we have used $A \sim 1 +2\kappa$ in the weak lensing regime, with $\kappa$ the lensing convergence; the second bracket comes from $(\chi_g/\chi)^2d\chi_g/d\chi$, and we have adopted the distant observer approximation, also restricted ourselves to sub-horizon scales \cite{Kaiser87, MatS96}.
By defining the intrinsic galaxy density to be
\begin{equation}
n_g(\chi_g,\boldsymbol{\theta_g})=\int \epsilon(e^{-\bar{\tau}(\chi_g,\lambda_{\rm obs})}f_g^{\lambda_g})\Phi_g(\chi_g,\boldsymbol{\theta_g};f_g^{\lambda_g})d f_g^{\lambda_g},
\end{equation}
we get
\begin{equation}
n(\chi,\boldsymbol{\theta})=n_g(\chi_g,\boldsymbol{\theta_g})\left[ 1 - 2\kappa - \frac{(1+z_g)}{H(z_g)} \frac{\partial v_{\parallel}}{\partial \chi_g}+2.5s_g(2\kappa - \delta \tau)\right]\label{eqn:nobs},
\end{equation}
with $s_g$ defined by
\begin{equation}
s_g(\chi_g)\equiv \frac{1}{2.5\bar{n}_g(\chi_g)}\int \frac{d\epsilon}{df^{\lambda}}\bigg|_{f^{\lambda}=e^{-\bar{\tau}}f_g^{\lambda_g}} e^{-\bar{\tau}}f_g^{\lambda_g}\bar{\Phi}_g(\chi_g;f_g^{\lambda_g}) df_g^{\lambda_g},
\end{equation}
where quantities with overbar represent their mean values. Up to the first order of perturbations, we find $s_g(\chi_g)=s(\chi)$, with
\begin{equation}
s(\chi)\equiv \frac{1}{2.5\bar{n}(\chi)}\int \frac{d\epsilon}{df^{\lambda}} f^{\lambda}\bar{\Phi}(\chi;f^{\lambda}) df^{\lambda}\label{eqn:sdef},
\end{equation}
which practically can be obtained from observations. When $\epsilon(f^{\lambda})$ is a step function, Eqn~(\ref{eqn:sdef}) reduces to Eqn~(\ref{eqn:ngslope}). Note $m=-2.5\log_{10} f^{\lambda}$ with a constant offset.

Finally, with the definition for the observed galaxy overdensity $\delta_{\rm obs}(\chi,\boldsymbol{\theta})\equiv(n-\bar{n})/\bar{n}$ and that for the intrinsic overdensity $\delta_g(\chi_g,\boldsymbol{\theta_g})\equiv(n_g-\bar{n}_g)/\bar{n}_g$, we obtain our results from Eqn~(\ref{eqn:nobs})
\begin{equation}
\delta_{\rm obs}=\delta_g-\frac{(1+z)}{H(z)}\frac{\partial v_{\parallel}}{\partial \chi}+[5s(z)-2]\kappa-2.5s(z)\delta\tau,
\end{equation}
where instead of $(\chi_g,\boldsymbol{\theta_g})$, the quantities on the right hand side are evaluated at $(\chi,\boldsymbol{\theta})$, same as $\delta_{\rm obs}$, which is accurate to first order in perturbations \cite{HuiGL07a}.

\onecolumngrid
\begin{center}
  {\bf APPENDIX B}
\end{center}

In this Appendix, we give the expressions for the individual $\xi_{\rm ab}$s with distinct combinations of ``ab", without keeping everything to the lowest order of $|\chi_i-\bar{\chi}|$ as done in \S \ref{subsec:xiobs}. The calculations of these terms are straightforward, 
so we just present the results here for reference.

The intrinsic galaxy correlation function is given by
\begin{equation}
\xi_{gg}(1;2)=\int\frac{d^3k}{(2\pi)^3}e^{i\bf{k}\cdot(\bf{x_1-x_2})}P_{gg}(k),
\end{equation}
where $P_{gg}(k)$ is the power spectrum for the galaxies at $z_1$ and $z_2$, with $z_i$ the redshift corresponding to $\chi_i$, and we do not explicitly specify this redshift dependence, similar for $P_{gm}(k)$ and $P_{mm}(k)$ below. With the plane-parallel approximation and the sub-horizon limit of the continuity equation, the corrections from the peculiar velocities are given by
\begin{eqnarray}
\xi_{gv}(1;2)&=& f_D (2) \int\frac{d^3k}{(2\pi)^3}e^{i\bf{k}\cdot(\bf{x_1-x_2})}(\hat{k}\cdot\hat{z})^2 P_{gm}(k), \\
\xi_{vv}(1;2)&=& f_D (1) f_D (2) \int\frac{d^3k}{(2\pi)^3}e^{i\bf{k}\cdot(\bf{x_1-x_2})}(\hat{k}\cdot\hat{z})^4 P_{mm}(k),
\end{eqnarray}
where $f_D  (i) =f_D (z_i)$, and
$f_D \equiv d\ln{D}/d\ln{a}$.
By using the Limber approximation and the Poisson equation, magnification bias in addition gives the following corrections
\begin{eqnarray}
\xi_{g\mu}(1;2) &=&\frac{3}{2}\Omega_m H_0^2(5s_2-2)(1+z_1)\Theta(\chi_2-\chi_1)\chi_1(1-\frac{\chi_1}{\chi_2})\times\nonumber  \\&& \int\frac{d^2k_{\perp}}{(2\pi)^2}e^{i\bf{k_{\perp}}\cdot \chi_1(\boldsymbol{\theta_1}-\boldsymbol{\theta_2})} P_{gm}(k_{\perp},z_1), \\
\xi_{\mu\mu}(1;2) &=& \left(\frac{3}{2}\Omega_m H_0^2\right)^2(5s_1-2)(5s_2-2)\int_0^{\min(\chi_1, \chi_2)} d\chi(1+z)^2\chi^2 \times \nonumber \\&& (1-\frac{\chi}{\chi_1})(1-\frac{\chi}{\chi_2})\int\frac{d^2k_{\perp}}{(2\pi)^2}e^{i\bf{k_{\perp}}\cdot \chi(\boldsymbol{\theta_1}-\boldsymbol{\theta_2})} P_{mm}(k_{\perp},z),\\
\xi_{v\mu}(1;2)&=&0,
\end{eqnarray}
where $s_i$ is $s(z)$ for the galaxies at redshift $z_i$, and $\Theta$ is the heaviside step function. Finally, with dust extinction, $\xi_{\rm obs}$ has the following extra corrections,
\begin{eqnarray}
\xi_{ge}(1;2)&=&-2.5s_2(1+z_1)^{-1}\bar{\rho}_d(z_1)f(z_1,\lambda_{\rm obs})\Theta(\chi_2-\chi_1) \int\frac{d^2k_{\perp}}{(2\pi)^2}e^{i\bf{k_{\perp}}\cdot \chi_1(\boldsymbol{\theta_1}-\boldsymbol{\theta_2})} P_{gd}(k_{\perp},z_1),\\
\xi_{ve}(1;2)&=&0,
\end{eqnarray}
where, again, we have used the Limber approximation. Since both $\xi_{\mu e}$ and $\xi_{ee}$ are negligible, we do not give their expressions here.

\end{document}